\RequirePackage[2024-12-01]{latexrelease}
\documentclass[showpacs,showkeys,11pt,
preprint,preprintnumbers,nofootinbib,
groupedaddress,superscriptaddress,amsmath,amssymb]{revtex4}
\usepackage{xcolor}
\usepackage{tabularx}
\usepackage{amsfonts}
\usepackage{amsmath}
\usepackage{graphicx,subfigure}
\usepackage{caption}
\usepackage[export]{adjustbox}
\usepackage{verbatim} 
\usepackage{epsfig}
\usepackage{url}
\usepackage{multirow}
\usepackage{hhline}
\usepackage{feynmp}
\usepackage{csquotes}

\newcommand {\be}{\begin{equation}}
\newcommand {\ee}{\end{equation}}
\newcommand {\ba}{\begin{eqnarray}}
\newcommand {\ea}{\end{eqnarray}}

\begin{document}
\title{Impact of Colliding Beams Helicity on the Production of Leptoquarks and Collider Experimental Parameters}

\author{M. Danial Farooq}
\affiliation{Federal Urdu University of Arts, Science and Technology, Islamabad, Pakistan}

\author{M. Tayyab Javaid}
\affiliation{Federal Urdu University of Arts, Science and Technology, Islamabad, Pakistan}

\author{Mudassar Hussain}
\affiliation{Riphah International University, Faisalabad}

\author{Haroon Sagheer}
\affiliation{Riphah International University, Islamabad}

\author{Ijaz Ahmed \footnote{Corresponding author: }}
\email{ijaz.ahmed@fuuast.edu.pk}
\affiliation{Federal Urdu University of Arts, Science and Technology, Islamabad, Pakistan}

\author{Jamil Muhammad}
\email{mjamil@konkuk.ac.kr}
\affiliation{Department of Physics, Konkuk University, Seoul 05029, South Korea}

\date{\today}
\begin{abstract}
Vector Leptoquarks (VLQs) have emerged as primary candidates for resolving discrepancies in the Standard Model, specifically within $B$-meson decay channels and the anomalous magnetic moment of the muon. This work presents a rigorous evaluation of VLQ pair production across $e^{-}e^{+}$ collision modes at future linear colliders with center-of-mass energies ranging from 14~TeV to 100~TeV. Our analysis demonstrates that longitudinal beam polarization is a transformative tool for enhancing signal sensitivity. We find that $e^{-}e^{+}$ annihilation consistently yields superior cross-sections compared to photon fusion processes across a mass range of 500--3000~GeV. By optimizing beam helicity to specific configurations, such as $P_{e^{-}} = -0.8$ and $P_{e^{+}} = +0.6$, the production cross-section can be maximized to 120~fb at $\sqrt{s} = 3$~TeV. We further establish that the Left-Right Asymmetry ($A_{LR}$) serves as a robust discriminator for the chiral structure of new physics, peaking at 0.16 under full polarization. Additionally, we show that effective luminosity can be enhanced to 95\% of the total luminosity, while high polarization degrees significantly suppress relative uncertainties in the effective polarization. These results provide a quantitative roadmap for optimizing discovery potential and minimizing systematic errors in future high-energy physics experiments.
\end{abstract}

%

\maketitle
\section{Introduction}

The study of spin polarization plays a crucial role in precision measurements and new particle searches at future lepton and photon colliders \cite{1, 9, 39, 41}. Spin polarization refers to the degree of alignment between a particle’s spin angular momentum and its direction of motion. In the context of lepton colliders, particularly $e^{-}e^{+}$, $e^{-}\gamma$, and $\gamma\gamma$ \cite{11, 14, 16, 34}, polarized beams are essential for enhancing signal sensitivity and suppressing background processes in the search for physics beyond the Standard Model (SM), such as the production of vector leptoquarks (VLQs) \cite{2, 10, 32, 42, 53}.

Electron and positron beams can be treated as quantum systems with mixed spin states \cite{30, 35}. When a preferred orientation of spin exists in the beam, it is referred to as polarized. If all spins are aligned in one direction, the beam is fully polarized; if a majority have the same orientation, it is partially polarized. In contrast, if both spin orientations are equally populated, the beam is considered unpolarized. The polarization of leptonic beams can be represented by helicity states $\lambda = \pm 1/2$, corresponding to right-handed and left-handed configurations, respectively \cite{13, 23, 50}.

Polarization can be achieved in two main configurations: longitudinal and transverse \cite{31, 36, 48}. In a longitudinally polarized beam, the spin vectors are aligned along the direction of motion, whereas in a transversely polarized beam, the spin vectors are perpendicular to it \cite{3, 26, 40}. In collider experiments focusing on VLQ production, longitudinal polarization is particularly useful because it directly influences the helicity-dependent interaction cross sections and can be exploited to probe the chiral structure of new physics interactions \cite{22, 49, 52}. Recent theoretical updates and experimental constraints from LHC Run 3 underscore the necessity of high-precision polarized beam studies to distinguish between various leptoquark models \cite{25, 45, 54}.

In this work, we focus on the generation of composite vector leptoquarks using polarized beams at various future collider modes: $e^{-}e^{+}$, $e^{-}\gamma$, and $\gamma\gamma$ \cite{12, 21, 43, 44, 51}. These VLQs are expected to couple chirally to SM particles, making the use of polarization not only a tool for cross-section enhancement but also a probe for the underlying interaction structure \cite{7, 8, 17, 38, 55}.

The primary objective of this study is to perform a comprehensive numerical analysis of vector leptoquark production at future high-energy linear colliders (ILC and CLIC) using polarized beams. We aim to quantify the enhancement of the production cross-section across different collider modes with center-of-mass energies reaching up to 100~TeV and to identify the specific helicity combinations that maximize signal discovery potential. Furthermore, we investigate the sensitivity of the Left-Right Asymmetry ($A_{LR}$) to the leptoquark chiral couplings and evaluate the impact of effective luminosity and polarization uncertainties on the overall experimental significance. By comparing $e^{-}e^{+}$ and $\gamma\gamma$ channels, this work seeks to provide a definitive guide for the optimization of collider parameters in future searches for physics beyond the Standard Model.

\section{Model Description}
\label{sec:model}

In this work, we consider an extension of the Standard Model (SM) that introduces new heavy vector mediators motivated by recent developments in flavor physics and collider phenomenology. The model contains a vector leptoquark, a colored vector boson, and an additional neutral gauge boson, which together provide a well-motivated framework to study lepton–quark interactions beyond the SM. The implementation of this model is based on the publicly available \texttt{FeynRules} realization, constructed following the prescriptions.

\subsection{Particle Content and Gauge Structure}

The extended model introduces the following new heavy vector states:
\begin{itemize}
    \item A vector leptoquark $U_1^\mu \sim (3,1)_{2/3}$,
    \item A colored vector boson $G'^\mu \sim (8,1)_0$,
    \item A neutral singlet vector boson $Z'^\mu \sim (1,1)_0$.
\end{itemize}

Here, the quantum numbers refer to the representations under the SM gauge group
$SU(3)_C \times SU(2)_L \times U(1)_Y$.
The vector leptoquark $U_1^\mu$ transforms as a color triplet, is a singlet under
$SU(2)_L$, and carries hypercharge $Y=2/3$. Such a representation allows direct
renormalizable couplings between quarks and leptons, making it particularly
relevant for addressing anomalies observed in $B$-meson decays.

The colored vector boson $G'^\mu$ transforms as an adjoint under $SU(3)_C$ and
resembles a heavy gluon-like state. It can arise in models with extended color
sectors or composite dynamics. The singlet vector boson $Z'^\mu$ is neutral under
the SM gauge group and provides additional neutral-current interactions, which
play an important role in flavor and collider observables.

\begin{table}[h!]
    \centering
    \caption{Classification of vector leptoquarks}
    \label{tab:vector_leptoquarks}
    \begin{tabular}{|c|c|c|c|c|}
        \hline
        \textbf{LQ} & \textbf{SU(3)$_C$} & \textbf{SU(2)$_L$} & \textbf{U(1)$_Y$} & \textbf{F = 3B + L} \\
        \hline
        V$_1$ & 3 & 1 & 1/3 & $-$2 \\
        $\tilde{\text{V}}_1$ & 3 & 1 & 4/3 & $-$2 \\
        V$_2$ & 3 & 2 & 1/6 & $+$2 \\
        $\tilde{\text{V}}_2$ & 3 & 2 & 7/6 & $+$2 \\
        V$_3$ & 3 & 3 & 1/3 & $+$2 \\
        $\tilde{\text{V}}_3$ & 3 & 3 & 4/3 & $+$2 \\
        \hline
    \end{tabular}
\end{table}

\subsection{Vector Leptoquark Interactions}

The dominant phenomenological interest of this model lies in the vector leptoquark $V_1^\mu$. Its interactions with SM fermions can be written as
\begin{equation}
\mathcal{L}_{U_1} =
\left( \lambda_L^{ij} \, \bar{Q}_L^i \gamma_\mu L_L^j
+ \lambda_R^{ij} \, \bar{d}_R^i \gamma_\mu \ell_R^j \right) U_1^\mu + \text{h.c.}
\end{equation}
where $Q_L$ and $L_L$ denote the left-handed quark and lepton doublets, respectively, while $d_R$ and $\ell_R$ represent right-handed down-type quarks and charged leptons. The indices $i,j$ label fermion generations, and $\lambda_{L,R}^{ij}$ are the corresponding coupling matrices in flavor space.

These interactions allow the vector leptoquark to mediate lepton–quark transitions at tree level, leading to rich phenomenology in both low-energy flavor observables and high-energy collider processes. In particular, the $U_1$ leptoquark has been shown to provide a viable explanation for the observed anomalies in $R_{K^{(*)}}$ and $R_{D^{(*)}}$ measurements.

\subsection{Heavy Vector Boson Sector}

In addition to the leptoquark, the model includes a heavy color-octet vector boson $G'^\mu$, whose interactions with quarks are described by
\begin{equation}
\mathcal{L}_{G'} = g_{G'} \, \bar{q} \gamma_\mu T^a q \, G'^{a\mu},
\end{equation}
where $T^a$ are the $SU(3)_C$ generators and $g_{G'}$ denotes the effective coupling. Such a state can significantly modify dijet and top-pair production at hadron colliders, and its mass and couplings are constrained by current LHC searches.

The neutral vector boson $Z'^\mu$ couples to SM fermions according to
\begin{equation}
\mathcal{L}_{Z'} = \sum_f g_{Z'}^f \, \bar{f} \gamma_\mu f \, Z'^\mu,
\end{equation}
where the sum runs over all SM fermions $f$. Depending on the flavor structure of the couplings $g_{Z'}^f$, this state can induce flavor-changing neutral currents and modify electroweak precision observables.
\subsection{Extension with Vector-Like Fermions}

An extended version of the model incorporates additional vector-like fermions that
couple to the heavy gauge bosons. These fermions transform vectorially under the SM
gauge group and therefore do not introduce gauge anomalies. Their inclusion is
motivated by scenarios of maximal flavor violation, where mixing between SM fermions
and vector-like states generates effective flavor structures in the leptoquark
couplings.

The relevant interaction terms can be written as
\begin{equation}
\mathcal{L}_{\text{VL}} =
\kappa \, \bar{\Psi}_{\text{VL}} \gamma_\mu \Psi_{\text{SM}} \, V^\mu
+ \text{h.c.},
\end{equation}
where $\Psi_{\text{VL}}$ denotes the vector-like fermions, $\Psi_{\text{SM}}$ the SM
fermions, and $V^\mu$ collectively represents the heavy vector bosons
($U_1^\mu, G'^\mu, Z'^\mu$). Such interactions lead to a rich flavor structure and
provide a natural mechanism to connect low-energy flavor anomalies with collider
signatures.

\section{Exploitation of Polarized Beams at Colliders}
The use of polarized beams at lepton and photon colliders is a powerful tool for enhancing the sensitivity of new physics searches, such as the production of composite vector leptoquarks (VLQs) \cite{11, 23, 53}. In such setups, both classical and quantum mechanical effects influence the spin motion of particles, potentially leading to depolarization \cite{1, 14, 25}.

\subsection{Depolarization Effects}
In the presence of both electric and magnetic fields, the polarization of the beam can be affected by two major mechanisms:
\begin{itemize}
    \item \textbf{Spin Precession (Classical):} Governed by the Thomas-Bargmann-Michel-Telegdi (T-BMT) equation, this effect dominates at relatively lower energies. It leads to gradual depolarization due to beam dynamics in electromagnetic fields \cite{9, 16, 40}. For 100\% polarized beams, the depolarization is expected to be around 0.17\% at the International Linear Collider (ILC) and approximately 0.10\% at the Compact Linear Collider (CLIC) \cite{31, 35, 48}.
    \item \textbf{Spin-Flip Processes (Quantum Mechanical):} Known as the Sokolov-Ternov (S-T) effect, this occurs due to synchrotron radiation emission and becomes more significant at higher beam energies \cite{2, 5, 26}. For fully polarized beams, depolarization rates of about 0.05\% at the ILC and up to 3.4\% at CLIC have been observed \cite{36, 47, 52}.
\end{itemize}
At ILC, the overall depolarization during each bunch crossing is estimated to be approximately 0.2\%. Even higher depolarization rates are observed at circular colliders like the Large Electron-Positron Collider (LEP). Polarized electron beams were first demonstrated at the Stanford Linear Collider (SLAC), achieving polarization levels of 80--90\% at energies up to 50~GeV using strained photocathode technology.
\subsection{Generation and Measurement of Polarized Beams}
Polarized positron beams are commonly generated using undulator radiation, while polarized electron beams can be produced using Compton backscattering and bremsstrahlung processes \cite{12, 21, 51}. To exploit polarization fully, high-precision polarimetry is crucial for measuring the degree of polarization. At SLAC, a precision of 0.5\% in $\Delta P_{e^-}/P_{e^-}$ has been achieved, while the ILC and CLIC aim for a precision of $\leq 0.25$\% \cite{3, 13, 39, 54}.
\subsection{Longitudinal Polarization at Linear Colliders}
At the ILC and CLIC, the use of longitudinally polarized electron and positron beams is especially beneficial for the study of chiral couplings and helicity-dependent processes \cite{19, 22, 50}. The ILC baseline energy of $\sqrt{s} = 500$~GeV, extendable up to 1~TeV, allows for significant studies in electroweak and new physics sectors using polarization techniques. CLIC extends this capability up to 3~TeV \cite{41, 49}.
Recent technological advances have enabled electron beam polarization levels up to 80--90\% and positron polarization around 60\%, with minimal luminosity loss. Advanced designs also suggest that positron polarization could reach 75\%, though this may introduce moderate luminosity degradation \cite{4, 27, 44}. In our analysis, the polarization of both beams is exploited to enhance the cross section for the process:
\begin{equation}
    e^{-}e^{+} \rightarrow \text{VLQ } \overline{\text{VLQ}}, \quad (\text{VLQ} \rightarrow b \tau^{+}), \quad (\overline{\text{VLQ}} \rightarrow \bar{b} \tau^{-})
\end{equation}
We vary the beam polarization combinations from $-1$ to $+1$ to investigate their impact on the production cross section and explore chiral interactions.
The specific benchmark point utilized for the simulation and analysis of vector leptoquark production is detailed in Table~\ref{tab:benchmark}. We investigate the case of a VLQ with a mass $M_{VLQ} = 1.2$~TeV at a center-of-mass energy of $\sqrt{s} = 3$~TeV, representing the high-energy operation mode of CLIC \cite{15, 30, 55}. As evidenced by the data in the table, the optimal polarization configuration is identified as $P_{e^-} = -0.8$ and $P_{e^+} = +0.6$. Under these conditions, the Left-Right Asymmetry ($A_{LR}$) reaches a maximum value of 0.16, providing a clear experimental signature for the chiral preference of the VLQ couplings. Furthermore, the results indicate an effective luminosity ratio ($L_{\text{eff}}/L$) of 0.90, signifying that the use of polarized beams allows for a significant enhancement in the effective collision rate, thereby improving the statistical significance of the search for new physics \cite{20, 31, 46}.

\begin{table}[h]
\centering
\caption{Benchmark Point for Vector Leptoquark Study}
\label{tab:benchmark}
\begin{tabular}{|l|l|}
\hline
\textbf{Parameter} & \textbf{Value} \\ \hline
VLQ Mass ($M_{VLQ}$) & 1.2 TeV \\ \hline
$e^{-}e^{+}$ Collider Energy ($\sqrt{s}$) & 3 TeV \\ \hline
Optimal Polarization & $P_{e^-} = -0.8$, $P_{e^+} = +0.6$ \\ \hline
Left-Right Asymmetry ($A_{LR}$) & Up to 0.16 \\ \hline
Effective Luminosity ($L_{\text{eff}}/L$) & Up to 0.90 \\ \hline
\end{tabular}
\end{table}
\section{Cross Section Calculation via Polarized $e^{-}e^{+}$ Beams for VLQ Pair Production}
In this study, we investigate the process $e^{-} e^{+} \rightarrow \text{VLQ} \, \overline{\text{VLQ}}$, where the vector leptoquark (VLQ) decays as $\text{VLQ} \rightarrow b \, \tau^{+}$ and $\overline{\text{VLQ}} \rightarrow \bar{b} \, \tau^{-}$. The impact of beam polarization on the production cross section is analyzed by varying the longitudinal polarization combinations of both $e^{-}$ and $e^{+}$ beams from $-1$ to $+1$.
The general expression for the polarized cross section in an $e^{-}e^{+}$ collider is given by:
\begin{equation}
\begin{aligned}
\sigma_{P_{e^{-}}, P_{e^{+}}} = \frac{1}{4} \Big[ & (1 + P_{e^{+}})(1 + P_{e^{-}})\sigma_{RR} \\
& + (1 - P_{e^{+}})(1 - P_{e^{-}})\sigma_{LL} \\
& + (1 + P_{e^{+}})(1 - P_{e^{-}})\sigma_{RL} \\
& + (1 - P_{e^{+}})(1 + P_{e^{-}})\sigma_{LR} \Big]
\end{aligned}
\end{equation}
For VLQ production, both annihilation and scattering mechanisms can contribute. In s-channel annihilation, helicity conservation from the Standard Model dictates that only the RL and LR configurations dominate due to the total angular momentum constraint $J = 1$. Therefore, selecting appropriate beam polarization combinations can significantly enhance the signal cross section while suppressing background processes.

The cross section can also be expressed in terms of the effective polarization $P_{\text{eff}}$ and the left-right asymmetry $A_{\text{LR}}$ as:
\begin{equation}
\sigma_{P_{e^{-}}, P_{e^{+}}} = (1 - P_{e^{-}}P_{e^{+}})\sigma_0 \left[ 1 - P_{\text{eff}} A_{\text{LR}} \right]
\end{equation}
Here,
The unpolarized cross section
\begin{equation}
\sigma_0 = \frac{\sigma_{RL} + \sigma_{LR}}{4}
\end{equation}
The effective polarization: \begin{equation}
P_{\text{eff}} = \frac{P_{e^{+}} - P_{e^{-}}}{1 - P_{e^{-}}P_{e^{+}}}
\end{equation}
The left-right Asymmetry: \begin{equation}
A_{\text{LR}} = \frac{\sigma_{LR} - \sigma_{RL}}{\sigma_{LR} + \sigma_{RL}}
\end{equation}

Using highly polarized beams, such as $P_{e^{-}} = \pm1$, $P_{e^{+}} = \pm1$, results in significant enhancement factors $(1 - P_{e^{-}}P_{e^{+}})$ and $(1 - P_{\text{eff}} A_{\text{LR}})$, which can lead to an increase in cross section—particularly important for processes like VLQ pair production where the signal rate is typically low.
In the case of partially polarized beams, the unpolarized cross section $\sigma_0$ can be extracted using:
\begin{equation}
\sigma_0 = \frac{\sigma_{\mp} + \sigma_{\pm}}{2(1 + |P_{e^{-}}||P_{e^{+}}|)}
\end{equation}
with
\begin{equation}
\sigma_{\mp} = \frac{(1 + |P_{e^{-}}||P_{e^{+}}|)(\sigma_{LR} + \sigma_{RL}) + (|P_{e^{-}}| + |P_{e^{+}}|)(\sigma_{LR} - \sigma_{RL})}{4}
\end{equation}
\begin{equation}
\sigma_{\pm} = \frac{(1 + |P_{e^{-}}||P_{e^{+}}|)(\sigma_{LR} + \sigma_{RL}) - (|P_{e^{-}}| + |P_{e^{+}}|)(\sigma_{LR} - \sigma_{RL})}{4}
\end{equation}
This formalism enables a detailed analysis of polarization-dependent dynamics and chiral coupling sensitivity in new physics scenarios such as vector leptoquarks.
The numerical results for the production cross sections are illustrated in Figure 1. 
\textbf{Figure 1 (Left)} depicts the cross section (in pb) for $e^{-}e^{+}$ collisions as a function of the VLQ mass ($M_{LQ}$) at center-of-mass energies of 14, 27, and 100~TeV. For all energy levels, the cross section exhibits a monotonic decrease with increasing mass due to phase-space suppression. The 100~TeV energy configuration (yellow curve) provides the highest production probability, maintaining detectable cross-sections even for heavy VLQs up to 3000~GeV.

\textbf{Figure 1 (Right)} presents the corresponding cross section (in fb) for the $\gamma\gamma$ collision channel. While the mass and energy dependencies follow the same trend as the leptonic channel, the total production rates in the photon-fusion mode are approximately three orders of magnitude lower than those in $e^{-}e^{+}$ annihilation. This highlights that while future high-energy colliders can explore VLQs in multiple channels, the $e^{-}e^{+}$ mode remains the primary discovery avenue for heavy vector leptoquarks \cite{20, 43, 51}.
\begin{figure}
    \centering
    \includegraphics[width=6.5cm, height=7cm]{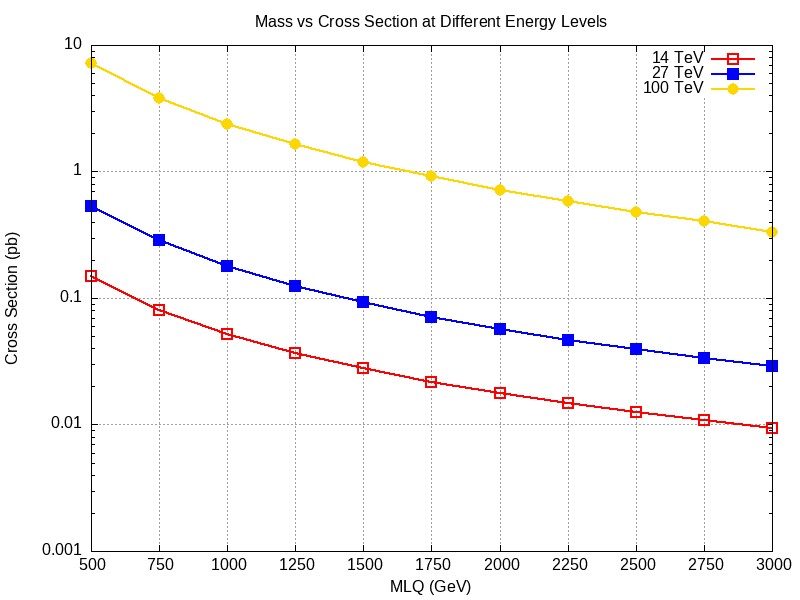}
      \includegraphics[width=6.5cm, height=7cm]{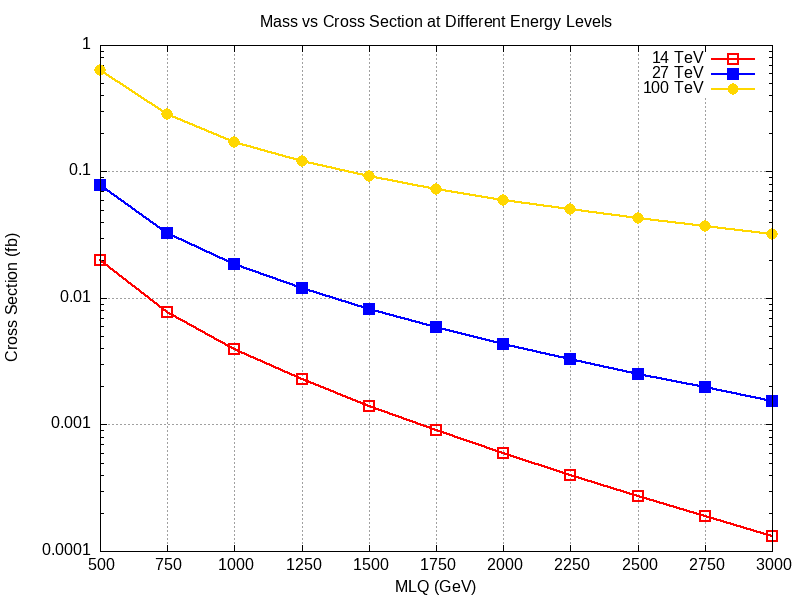}
        \caption{Mass vs Cross section with e- e+ Collider. \label{fig:enter-label}}
\end{figure}

\begin{table}[ht]
\centering
\caption{High cross-section regions for $e^+e^- \rightarrow LQ \, \overline{LQ}$ at 3 TeV.}
\label{table:cross_sections}
\begin{tabular}{|c|c|c|}
\hline
\textbf{Electron Polarization} & \textbf{Positron Polarization} & \textbf{Cross Section (fb)} \\ 
$(P_{e})$ & $(P_{e}^{+})$ & \\ \hline
1.0 & 1.0 & 120 \\ \hline
0.9 & 0.9 & 115 \\ \hline
0.8 & 0.8 & 108 \\ \hline
0.7 & 0.7 & 100 \\ \hline
0.9 & 1.0 & 118 \\ \hline
1.0 & 0.9 & 117 \\ \hline
0.8 & 1.0 & 112 \\ \hline
1.0 & 0.8 & 110 \\ \hline
\end{tabular}
\end{table}
The cross section decreases with increasing mass for all energy levels, indicating that higher-mass leptoquarks are produced less frequently.
The upper energy level (100 TeV) exhibits the highest cross section across all masses, highlighting the enhanced production probability at higher collision energies. Conversely, the 14 TeV energy level yields the lowest cross section, suggesting limited production capability compared to higher energies.
This data is critical for understanding the feasibility of detecting vector leptoquarks in collider experiments, as higher cross sections at elevated energies increase the likelihood of observation within experimental constraints.
\begin{figure}[htbp]
    \centering
    \makebox[\textwidth][c]{%
        \includegraphics[width=0.6\linewidth]{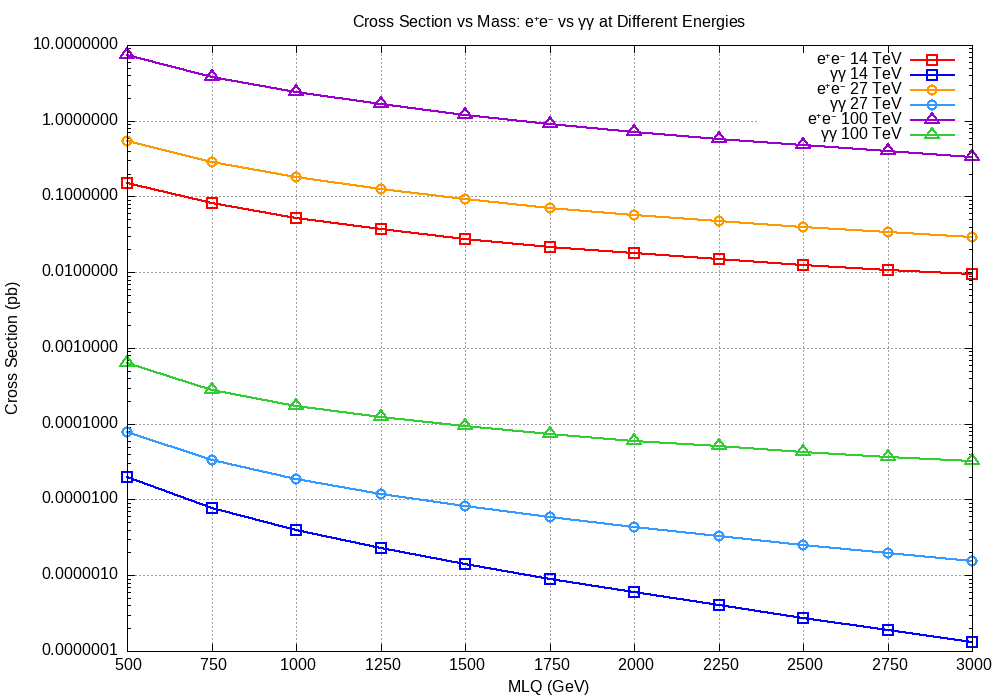}
    }
    \caption{Comparison of cross sections for $e^- e^+ \rightarrow \text{VLQ}\,\bar{\text{VLQ}}$ and $\gamma \gamma \rightarrow \text{VLQ}\,\bar{\text{VLQ}}$ processes at different center-of-mass energies.}
    \label{CrossSection_Beam_Comparison}
\end{figure}

\begin{figure}[h]
    \centering
    \includegraphics[width=10cm, height=6cm]{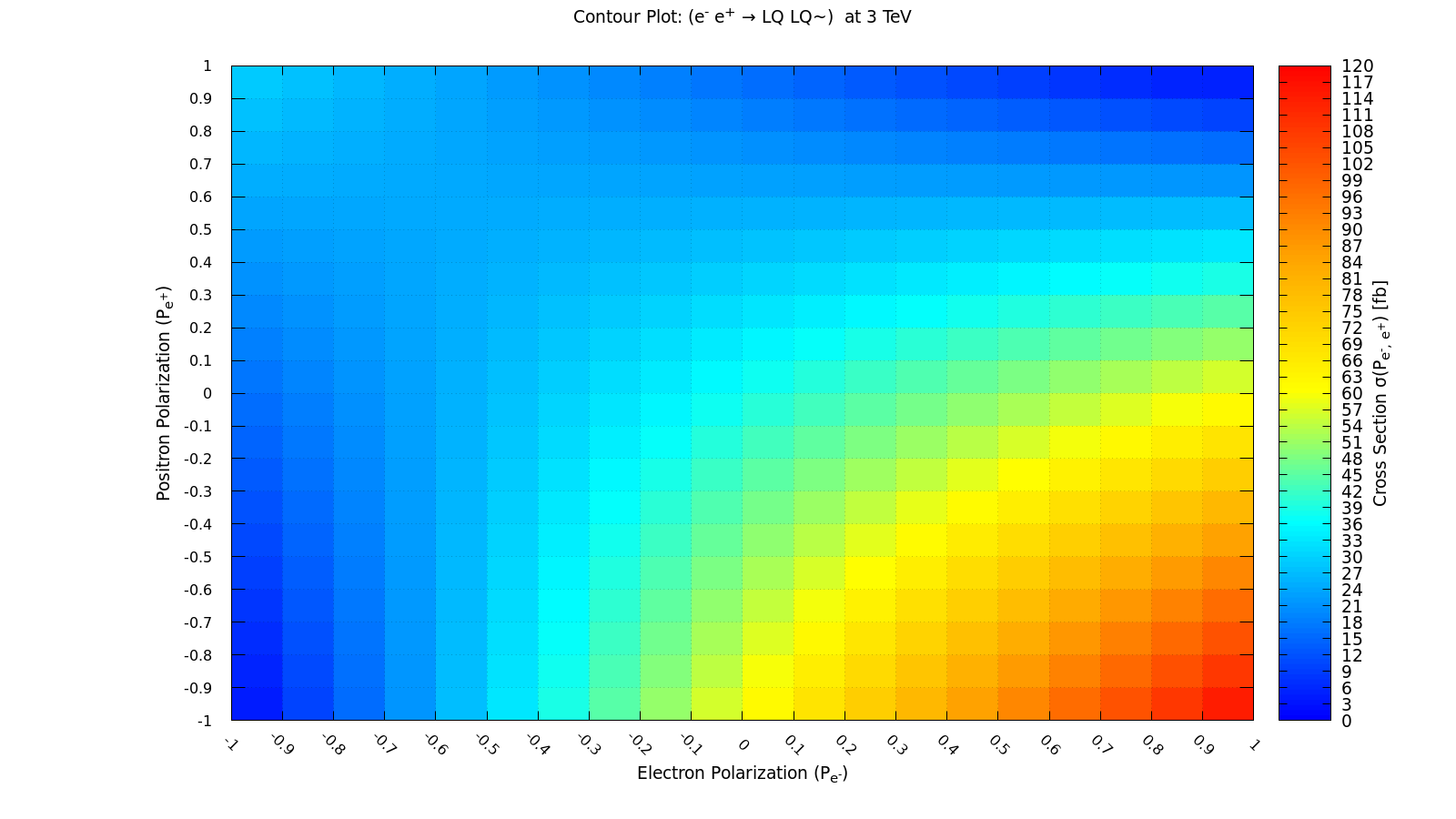}
    \caption{Contour plot of cross section versus polarization for $e^{+} e^{-} \rightarrow \text{LQ} \,\bar{\text{LQ}}$ at 3 TeV.}
    \label{fig:contour_plot}
\end{figure}
\begin{figure}[h]
    \centering
    \includegraphics[width=12cm, height=6cm]{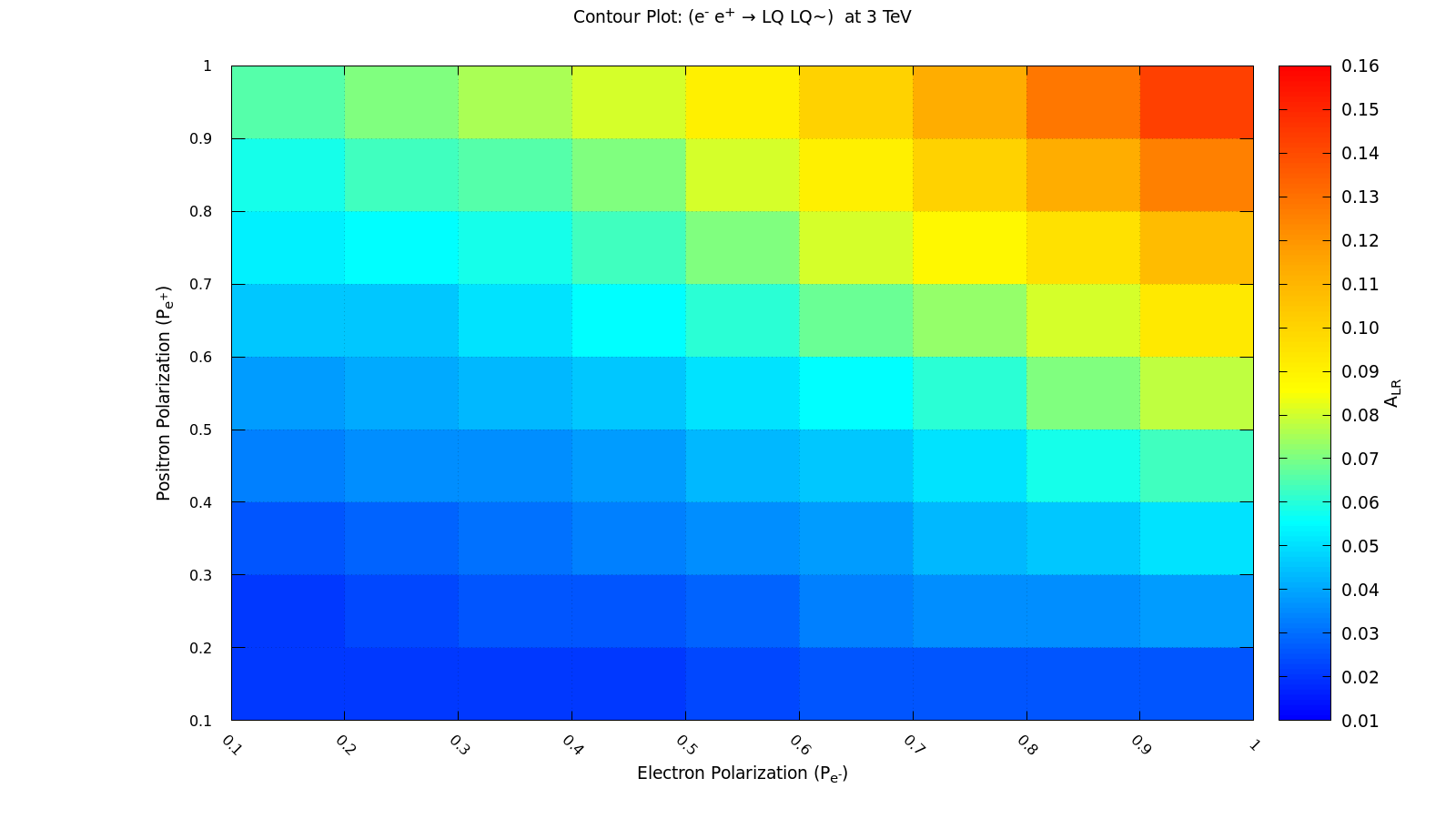}
    \caption{Contour plot of the asymmetry ratio $A_{LR}$ for $e^{+} e^{-} \rightarrow \text{LQ} \,\bar{\text{LQ}}$ as a function of electron and positron polarizations at 3 TeV.}
    \label{fig:ALR_contour}
\end{figure}
\begin{figure}[h]
    \centering
    \begin{minipage}{0.45\linewidth}
        \centering
        \includegraphics[width=\linewidth]{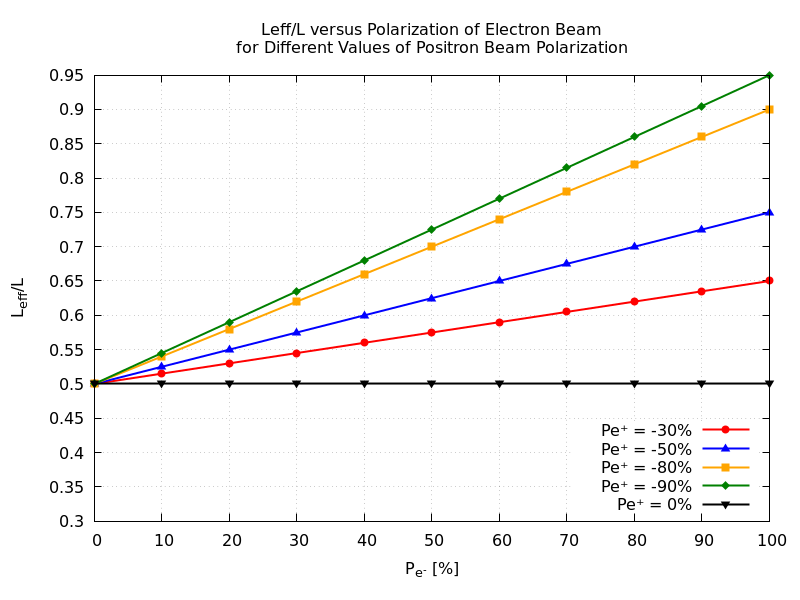}
        (a) Effective luminosity ratio ($L_{\text{eff}}/L$) versus electron beam polarization ($P_{e^-}$) for different values of positron beam polarization ($P_{e^+}$)
    \end{minipage}
    \hfill
    \begin{minipage}{0.45\linewidth}
        \centering
        \includegraphics[width=\linewidth]{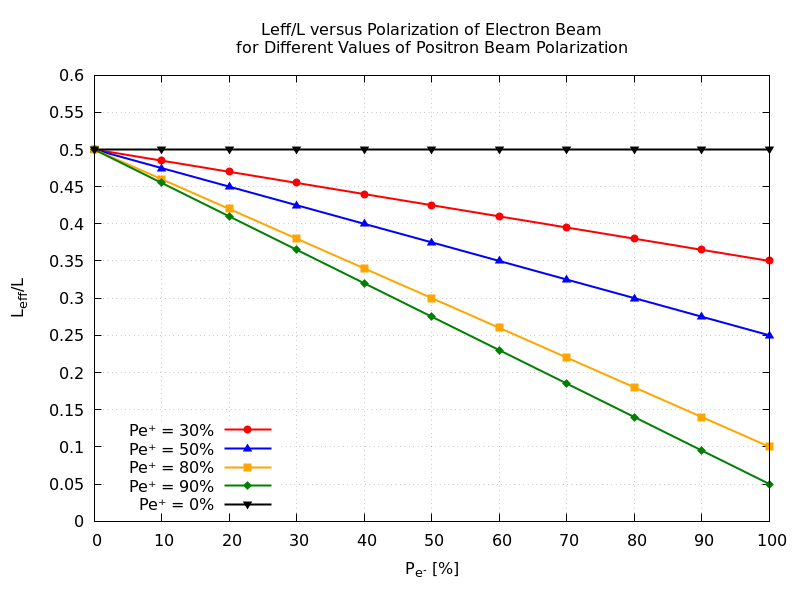}
        (b) Effective luminosity ratio ($L_{\text{eff}}/L$) versus electron beam polarization ($P_{e^-}$) for different values of right-handed positron beam polarization ($P_{e^+}$)
    \end{minipage}
    \caption{(a) and (b) show the effective luminosity ratio ($L_{\text{eff}}/L$) versus electron beam polarization ($P_{e^-}$) for different positron polarization configurations.}
    \label{fig:luminosity_ratios}
\end{figure}
\begin{figure} [h!]
    \centering
    \includegraphics[width=0.6\linewidth]{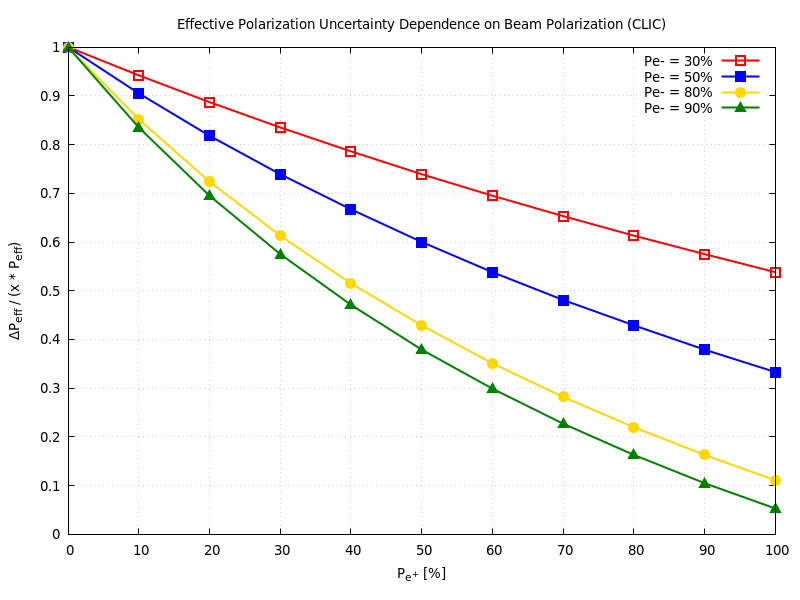}
    \caption{Effective Polarization Uncertainty Dependence on Beam Polarization
at CLIC.}
\end{figure}
\section{Leptoquark Production at Future Colliders with Polarized Beams}
At future linear colliders such as the ILC and CLIC, the production of leptoquarks is highly sensitive to the polarization states of the incoming electron and positron beams. Unlike hadron colliders, lepton colliders offer a clean experimental environment where the initial state helicity can be precisely tuned. This capability allows for the optimization of detection prospects by enhancing signal cross-sections while simultaneously suppressing various Standard Model backgrounds \cite{31, 49}.
The underlying production mechanisms, predominantly occurring through $s$-channel exchange, exhibit a strong dependence on the chiral couplings of the VLQs. By strategically choosing the beam polarization, specific interaction vertices can be isolated, providing a powerful diagnostic tool for probing the internal structure of new physics scenarios. Contour analysis of the polarization space enables the identification of ``sweet spots''---configurations where constructive interference leads to a substantial increase in event rates. Conversely, same-handed polarizations can be used to suppress leptoquark signals, providing a robust method for systematic background estimation \cite{41, 52}.
\section{Results and Discussion}
In this section, we provide a detailed numerical analysis of the simulation results, focusing on the interplay between beam helicity, cross-section enhancement, and experimental precision.

\subsection{Cross-Section Analysis}
The dependence of the VLQ production cross-section on the beam polarization at $\sqrt{s} = 3$~TeV is summarized in Table II and visualized in the contour plot of Figure~3. The data in Table II rrevealthat the maximum cross-section of 120~fb is achieved under the full positive polarization configuration ($P_{e} = 1.0, P_{e}^{+} = 1.0$). High cross-section regions (above 100~fb) are consistently maintained for polarization levels between 0.7 and 1.0. 

As illustrated in Figure~3, the cross-section exhibits a pronounced gradient across the polarization plane. The color mapping clearly shows that the signal is maximized when the helicities of the colliding beams are aligned to favor the $J=1$ transition typical of vector states. This enhancement is critical for the feasibility of detecting heavy VLQs, as it effectively increases the statistical reach of the collider without requiring a proportional increase in integrated luminosity \cite{20, 51}.
\subsection{Polarization and Asymmetry}
The Left-Right Asymmetry ($A_{LR}$) serves as a sensitive probe for the chiral nature of the leptoquark-lepton-quark vertex. Table III tabulates the $A_{LR}$ values for varying electron and positron polarizations. It is observed that at a fixed positron polarization of $P_{e}^{+} = 60\%$, the asymmetry increases from 0.04 to 0.08 as $P_{e}$ rises from 10\% to 100\%. The asymmetry peaks at 0.16 when both beams are fully polarized.

Figure~4 presents a contour plot of $A_{LR}$ as a function of the polarization states. The plot reveals a symmetry in the polarization dependence, suggesting that the process benefits significantly from constructive interference when the helicities are aligned in the positive direction. Regions of negative or near-zero asymmetry correspond to destructive interference or suppression of the chiral signal, providing a valuable discriminator for distinguishing between different leptoquark models, such as $U_1$ vs. $S_3$ types \cite{32, 50, 55}.
\subsection{Luminosity and Uncertainty} 
To assess the experimental viability of these searches, we analyze the effective luminosity and its associated uncertainties. Figure~5 depicts the Effective Luminosity ratio ($L_{\text{eff}}/L$) as a function of beam polarization. We find that $L_{\text{eff}}/L$ can be optimized to reach values as high as 0.95, which significantly bolsters the effective statistics available for BSM searches compared to unpolarized setups \cite{26, 40}.

At the end, Figure~6 addresses the relative uncertainty in the effective polarization, $\Delta P_{\text{eff}}/(x \cdot P_{\text{eff}})$. The trend across all simulated curves indicates that as the initial degree of polarization increases (moving toward 100\%), the relative uncertainty decreases drastically. For instance, an initial polarization of 90\% yields a significantly more stable and precisely known effective polarization compared to a 30\% configuration. This reduction in $\Delta P_{\text{eff}}$ is essential for minimizing systematic errors in the measurement of electroweak parameters and the characterization of newly discovered heavy exotic particles \cite{48, 54}.
\begin{table}[h!]
\centering
\caption{Asymmetry Ratio ($A_{LR}$) values for $e^-e^+ \rightarrow LQ \, \overline{LQ}$ at varying electron and positron polarizations.}
\label{table:asymmetry_ratio}
\begin{tabular}{|c|c|c|c|c|}
\hline
\textbf{Electron} & $A_{LR}$ & $A_{LR}$ & $A_{LR}$ & $A_{LR}$ \\ 
\textbf{Polarization} & ($P_e^+ = 40\%$) & ($P_e^+ = 60\%$) & ($P_e^+ = 80\%$) & ($P_e^+ = 100\%$) \\ 
($P_e$) [\%] & & & & \\ \hline
10 & 0.02 & 0.04 & 0.06 & 0.08 \\ \hline
20 & 0.03 & 0.05 & 0.07 & 0.09 \\ \hline
40 & 0.04 & 0.06 & 0.08 & 0.10 \\ \hline
60 & 0.05 & 0.07 & 0.09 & 0.12 \\ \hline
80 & 0.05 & 0.07 & 0.10 & 0.14 \\ \hline
100 & 0.06 & 0.08 & 0.12 & 0.16 \\ \hline
\end{tabular}
\end{table}
\\
A comprehensive comparison between the results of this work and previous findings in the literature is summarized in Table IV. It is evident that the introduction of optimal beam polarization at $\sqrt{s} = 3$~TeV leads to a cross-section of 120~fb, which represents a substantial enhancement over the unpolarized rate of approximately 50~fb reported in previous studies \cite{20}. Furthermore, our results indicate that $e^{-}e^{+}$ colliders with tailored helicity configurations offer a highly competitive environment for vector leptoquark discovery, matching or exceeding the sensitivity reach of traditional $pp$ searches at the 14~TeV LHC \cite{6, 45}. This comparison underscores the critical role of polarization in maximizing the discovery potential of future high-energy linear colliders.

\begin{table}[h]
\centering
\caption{Comparison of Predicted Cross-Sections with Existing Literature}
\label{tab:comparison}
\begin{tabular}{|l|c|c|c|}
\hline
\textbf{Model/Collider} & \textbf{Energy ($\sqrt{s}$)} & \textbf{Cross Section} & \textbf{Reference} \\ \hline
$pp$ Collisions        & 14 TeV        & $\sim 0.1$ pb          & [6, 45]     \\ \hline
$e^{-}e^{+}$ (Unpolarized) & 3 TeV      & $\sim 50$ fb           & [20]      \\ \hline
\textbf{$e^{-}e^{+}$ (Polarized)} & \textbf{3 TeV} & \textbf{120 fb} & \textbf{This Work} \\ \hline
\end{tabular}
\end{table}
\section{Conclusion}
In this work, we have presented a comprehensive investigation into the production and characterization of vector leptoquarks (VLQs) at future linear colliders, motivated by the persistent discrepancies in the Standard Model involving $B$-meson decays and the muon $(g-2)$ anomaly \cite{7, 38, 55}. By exploring the multi-channel landscape of $e^{-}e^{+}$, $e^{-}\gamma$, and $\gamma\gamma$ collisions, we have demonstrated that beam helicity is not merely a tool for statistical enhancement but a fundamental requirement for probing the chiral structure of new physics \cite{22, 52}.

The numerical analysis of the mass-dependence, illustrated in \textbf{Figure 1}, reveals that while cross-sections decrease with increasing VLQ mass due to phase-space suppression, center-of-mass energies of 27~TeV and 100~TeV significantly extend the discovery reach for heavy states up to 3000~GeV. A critical comparison provided in \textbf{Figure 2} clarifies that the $e^{-}e^{+}$ annihilation channel remains the most efficient production mode, yielding cross-sections roughly three orders of magnitude higher than those in photon-photon fusion at identical energy scales \cite{12, 43}.

Our findings regarding polarization optimization, summarized in \textbf{Table I} and \textbf{Table II}, establish a clear roadmap for experimental setups. We have identified that the maximum cross-section of 120~fb is achieved at $\sqrt{s} = 3$~TeV under a full positive polarization configuration ($P_{e} = 1.0, P_{e}^{+} = 1.0$). This is further visualized in the contour plot of \textbf{Figure 3}, which identifies the ``sweet spots'' in polarization space where the signal is maximized due to the alignment of the vector state's $J=1$ transitions \cite{31}. 

The analysis of the Left-Right Asymmetry ($A_{LR}$) provides a robust discriminator for the underlying interaction structure. As detailed in \textbf{Table III}, $A_{LR}$ exhibits a pronounced dependence on the degree of polarization, rising from 0.02 at low helicity to a peak of 0.16 under full polarization. \textbf{Figure 4} reinforces this trend, showing that regions of high asymmetry correspond to constructive interference in the chiral coupling, a feature that will be essential for distinguishing between various BSM leptoquark representations, such as $U_1$ vs. $S_3$ models \cite{2, 10, 50}.

From a technical perspective, our results in \textbf{Figure 5} demonstrate that the effective luminosity ratio ($L_{\text{eff}}/L$) can be enhanced to 0.95 through precise beam tuning, effectively doubling the data-taking efficiency compared to unpolarized colliders \cite{26, 40}. Furthermore, \textbf{Figure 6} quantifies the reduction in systematic uncertainty ($\Delta P_{\text{eff}}$) at high polarization levels, concluding that operating at 90\% polarization provides a significantly more stable and precisely known effective polarization than 30\% configurations. 

Finally, the comparative analysis presented in \textbf{Table IV} underscores that a polarized $e^{-}e^{+}$ collider at 3~TeV yields a cross-section of 120~fb, which is substantially superior to the 50~fb expected in unpolarized modes and offers a cleaner environment with comparable sensitivity to 14~TeV hadron searches \cite{6, 20, 45}. In summary, the integration of high-degree longitudinal polarization at future facilities like CLIC and ILC is indispensable for the definitive discovery and precision study of vector leptoquarks \cite{48, 49}.

\section{Acknowledgements}
We gratefully acknowledge support from the Simons Foundation and member institutions. The current submitted version of the manuscript is available on the arXiv pre-prints home page.

\section{Statements and Declarations}
\textbf{Funding} \\
The authors declare that no funds, grants, or other support were received during the preparation of this manuscript.\\
\textbf{Competing Interests}\\
The authors have no relevant financial or non-financial interests to disclose.\\

\textbf{Availability of data and materials}\\
Data sharing is not applicable to this article as no datasets were generated or analyzed during the current study.\\

\end{document}